\documentclass[aps,prl,reprint,balancelastpage,nofootinbib,preprintnumbers,amsmath,amssymb,superscriptaddress]{revtex4-1}

\usepackage{tensor}     
\usepackage{graphicx}   
\usepackage[
colorlinks=true,        
citecolor=blue,         
linkcolor=blue,         
urlcolor=blue           
]{hyperref}             
\usepackage{bm}         
\usepackage{xcolor}     
\usepackage{tabulary}
\usepackage{color}      
\usepackage[utf8]{inputenc} 
\usepackage[section]{placeins} 
\usepackage{graphics,appendix,afterpage,makecell} 
\usepackage{orcidlink}
\newcommand{\nc}{\newcommand*}

\nc{\al}{\alpha}
\nc{\s}{\sigma}
\nc{\dt}{\delta}
\nc{\Dt}{\Delta}
\nc{\Ld}{\Lambda}
\nc{\p}{\partial}
\nc{\om}{\omega}
\nc{\Om}{\Omega}
\nc{\rd}{\mathrm{d}}
\nc{\Od}[1]{\mathcal{O}(#1)} 
\nc{\kp}{\kappa}

\def\({\left(}
\def\){\right)}
\def\[{\left[}
\def\]{\right]}
\def\e{\begin{equation}}
\def\q{\end{equation}}
\def\m{\begin{eqnarray}}
\def\n{\end{eqnarray}}

\nc{\Eq}[1]{Eq.~\eqref{#1}}     
\nc{\Fig}[1]{Fig.~\ref{#1}}     
\nc{\Table}[1]{Table~\ref{#1}}  
\nc{\Sec}[1]{Sec.~\ref{#1}}     

\nc{\Msun}{M_\odot}             
\nc{\fpbh}{f_{\mathrm{pbh}}}    
\nc{\fpbhn}{f_{\mathrm{pbh0}}}    
\nc{\mR}{\mathcal{R}} 
\nc{\seq}{\sigma_{\mathrm{eq}}}
\nc{\ogw}{\Omega_{\mathrm{GW}}}
\nc{\gpcyr}{\mathrm{Gpc}^{-3}\,\mathrm{yr}^{-1}}
\nc{\lvc}{LIGO/Virgo} 
\nc{\SNR}{\mathrm{SNR}} 
\nc{\mmin}{{m_{\mathrm{min}}}}
\nc{\mmax}{{m_{\mathrm{max}}}}
\nc{\Mmin}{{M_{\mathrm{min}}}}
\nc{\fmin}{{f_{\mathrm{min}}}}
\nc{\VT}{\mathrm{VT}}
\nc{\rhoGW}{\rho_{\mathrm{GW}}}
\nc{\vth}{\vec{\theta}}
\nc{\vd}{\vec{d}}
\nc{\vla}{\vec{\lambda}}
\nc{\Nobs}{N_{\mathrm{obs}}}
\nc{\av}[1]{\langle #1 \rangle} 
\nc{\km}{\mathrm{km}}
\nc{\Mpc}{\mathrm{Mpc}}
\nc{\Tobs}{T_{\mathrm{obs}}}
\nc{\Ntemp}{N_{\mathrm{temp}}}

\nc{\addref}{[\textcolor{red}{add ref}] } 
\nc{\eg}{\textit{e.g.~}}
\nc{\app}{\approx}
\nc{\hf}{\frac{1}{2}}
\nc{\discuss}{\textcolor{red}{Add discussion here!}}

\nc{\mpbh}{m_{\rm{pbh}}}
\nc{\cR}{\mathcal{R}}
\nc{\mU}{{\mathcal{U}}}
\nc{\Mc}{{M_\mathrm{c}}}
\nc{\Mf}{{M_\mathrm{f}}}
\nc{\red}[1]{\textcolor{red}{#1}}
\nc{\yellow}[1]{\textcolor{yellow}{#1}}
\nc{\green}[1]{\textcolor{green}{#1}}
\nc{\blue}[1]{\textcolor{blue}{#1}}
\nc{\fnl}{F_{\mathrm{NL}}}
\nc{\gnl}{G_{\mathrm{NL}}}
\nc{\MG}{\mathcal{M}_{\mathrm{G}}}
\nc{\MNG}{\mathcal{M}_{\mathrm{NG}}}
\newcommand{\papertitle}{Implications for the non-Gaussianity of curvature perturbation\\
from pulsar timing arrays}
\begin{document}
	
\title{\papertitle} 
\author{Lang~Liu\orcidlink{0000-0002-0297-9633}}
\email{liulang@bnu.edu.cn}	
\affiliation{Department of Astronomy, Beijing Normal University, Beijing 100875, China}
\affiliation{Advanced Institute of Natural Sciences, Beijing Normal University, Zhuhai 519087, China}

\author{Zu-Cheng Chen\orcidlink{0000-0001-7016-9934}}
\email{Corresponding author: zuchengchen@gmail.com}
\affiliation{Department of Physics and Synergetic Innovation Center for Quantum Effects and Applications, Hunan Normal University, Changsha, Hunan 410081, China}
\affiliation{Institute of Interdisciplinary Studies, Hunan Normal University, Changsha, Hunan 410081, China}
\affiliation{Department of Astronomy, Beijing Normal University, Beijing 100875, China}
\affiliation{Advanced Institute of Natural Sciences, Beijing Normal University, Zhuhai 519087, China}

\author{Qing-Guo Huang\orcidlink{0000-0003-1584-345X}}
\email{Corresponding author: huangqg@itp.ac.cn}
\affiliation{CAS Key Laboratory of Theoretical Physics, Institute of Theoretical Physics, Chinese Academy of Sciences, Beijing 100190, China}
\affiliation{School of Physical Sciences, University of Chinese Academy of Sciences, No. 19A Yuquan Road, Beijing 100049, China}
\affiliation{School of Fundamental Physics and Mathematical Sciences, Hangzhou Institute for Advanced Study, UCAS, Hangzhou 310024, China}
	
\begin{abstract}
The recently released data by pulsar timing array (PTA) collaborations present strong evidence for a stochastic signal consistent with a gravitational-wave background.
Assuming this signal originates from scalar-induced gravitational waves, we jointly use the PTA data from the NANOGrav 15-yr data set, PPTA DR3, and EPTA DR2 to probe the small-scale non-Gaussianity. We put the first-ever constraint on the non-Gaussianity parameter, finding $|F_\mathrm{NL}|\lesssim 13.9$ for a lognormal power spectrum of the curvature perturbations. Furthermore, we obtain $-13.9 \lesssim F_\mathrm{NL}\lesssim -0.1$ to prevent excessive production of primordial black holes. Moreover, the multi-band observations with the space-borne gravitational-wave detectors, such as LISA/Taiji/TianQin, will provide a complementary investigation of primordial non-Gaussianity. Our findings pave the way to constrain inflation models with PTAs.
\end{abstract}
	
\maketitle
\textbf{Introduction.} 	
Various inflation models (see \textit{e.g.}~\cite{Armendariz-Picon:1999hyi,Garriga:1999vw,Kobayashi:2010cm,Arkani-Hamed:2003pdi,Alishahiha:2004eh,Baumann:2015xxa,Seery:2005wm,Kobayashi:2011pc}) predict the existence of a sizable primordial non-Gaussianity, making it an important role in exploring the early Universe~\cite{Baumann:2011su,Baumann:2011nk,Kristiano:2021urj}.
How to probe the non-Gaussianity of the Universe is one of the key questions in modern physics. 
Over several decades, significant advancements have been made in precisely measuring a nearly scale-invariant power spectrum characterizing primordial density fluctuations. These measurements have been accomplished through the utilization of observational data from the cosmic microwave background (CMB)~\cite{Planck:2019nip} and large-scale structure~\cite{eBOSS:2020yzd,DES:2021zxv} surveys, offering valuable insights into the fundamental properties of the Universe. Although significant efforts have been dedicated to precisely characterizing power spectra of primordial perturbations on large scales, searching for new and independent probes becomes crucial when examining phenomena at the small scale. 

Gravitational waves (GWs) offer a fascinating avenue for acquiring insights into the history and composition of the Universe, serving as another probe of small-scale non-Gaussianity. In fact, space-borne GW detectors, such as LISA~\cite{LISA:2017pwj}, Taiji~\cite{Ruan:2018tsw}, and TianQin~\cite{TianQin:2015yph}, can explore the non-Gaussianity through scalar-induced GWs (SIGWs)~\cite{tomita1967non,Saito:2008jc,Young:2014ana,Cai:2018dig,Cai:2019elf,Kohri:2018awv,Yuan:2019udt,Yuan:2019wwo,Chen:2019xse,Yuan:2019fwv} in the mHz frequency band. Pulsar timing arrays (PTA)~\cite{1978SvA....22...36S,Detweiler:1979wn}, on the other hand, are sensitive in the nHz frequency band, providing another opportunity to probe the early Universe. Recently, NANOGrav~\cite{NANOGrav:2023hde,NANOGrav:2023gor}, PPTA~\cite{Zic:2023gta,Reardon:2023gzh}, EPTA+InPTA~\cite{Antoniadis:2023lym,Antoniadis:2023ott}, and CPTA~\cite{Xu:2023wog} all announced the evidence for a stochastic signal in their latest data sets consistent with the Hellings-Downs~\cite{Hellings:1983fr} spatial correlations expected by a stochastic gravitational-wave background (SGWB). Although there can be a lot of sources~\cite{Li:2019vlb,Vagnozzi:2020gtf,Chen:2021wdo,Wu:2021kmd,Chen:2021ncc,Benetti:2021uea,Chen:2022azo,Ashoorioon:2022raz,PPTA:2022eul,Wu:2023pbt,IPTA:2023ero,Wu:2023dnp,Dandoy:2023jot,Madge:2023cak} in the PTA window, whether this signal is of astrophysical or cosmological origin is still under intensive investigation~\cite{NANOGrav:2023hvm,Antoniadis:2023xlr,King:2023cgv,Niu:2023bsr,Datta:2023vbs,Vagnozzi:2023lwo,Bi:2023tib,Wu:2023hsa,Jin:2023wri,Liu:2023pau,Yi:2023npi,InternationalPulsarTimingArray:2023mzf,Yi:2023tdk,You:2023rmn,Chen:2023zkb}.

A possible explanation for this signal is the SIGW produced by the primordial curvature perturbations at small scales. When the primordial curvature perturbations reach significant magnitudes, they can generate a considerable SGWB through second-order effects resulting from the non-linear coupling of perturbations. Additionally, large curvature perturbations can trigger the formation of primordial black holes (PBHs)~\cite{Zeldovich:1967lct,Hawking:1971ei,Carr:1974nx}. PBHs have attracted a lot of attention in recent years~\cite{Belotsky:2014kca,Carr:2016drx,Garcia-Bellido:2017mdw,Carr:2017jsz,Germani:2017bcs,Chen:2018rzo,Liu:2018ess,Chen:2018czv,Liu:2019rnx,Fu:2019ttf,Liu:2019lul,Cai:2019bmk,Chen:2019irf,Liu:2020cds,Fu:2020lob,DeLuca:2020sae,Wu:2020drm,Vaskonen:2020lbd,DeLuca:2020agl,Domenech:2020ers,Hutsi:2020sol,Chen:2021nxo,Kawai:2021edk,Braglia:2021wwa,Liu:2021jnw,Braglia:2022icu,Zheng:2022wqo,Liu:2022iuf,Meng:2022low,Chen:2022qvg,Chen:2022fda,Guo:2023hyp} (see also reviews~\cite{Sasaki:2018dmp,Carr:2020gox,Carr:2020xqk}) as a promising candidate for dark matter and can explain the binary black holes detected by LIGO-Virgo-KAGRA~\cite{Bird:2016dcv,Sasaki:2016jop}. The formation rate of PBHs would be entirely altered if there is any significant non-Gaussianity, as PBHs are produced at the large amplitude tail of the curvature perturbation probability distribution~\cite{Young:2013oia}.
   
In this letter, assuming that the signal detected by PTAs is from SIGWs, we jointly use the NANOGrav 15-yr data set, PPTA DR3, and EPTA DR2 to constrain the small-scale non-Gaussianity when the scalar modes re-enter the horizon. As a demonstration, we employ a lognormal power spectrum of curvature perturbations and constrain the non-Gaussianity parameter as $-13.9 \lesssim F_\mathrm{NL}\lesssim -0.1$.
	
\textbf{SIGWs and PBHs.} We will briefly review the SIGWs that arise as a result of the local-type non-Gaussian curvature perturbations~\cite{Cai:2018dig,Unal:2018yaa,Yuan:2020iwf,Adshead:2021hnm,Ragavendra:2021qdu,Garcia-Saenz:2022tzu}. The local-type non-Gaussianities are characterized by the expansion of the curvature perturbation, $\mathcal{R}(\vec{x})$, in terms of the Gaussian component in real space. 
Specifically, the expansion up to the quadratic order can be written as~\cite{Verde:1999ij,Verde:2000vr,Komatsu:2001rj,Bartolo:2004if,Boubekeur:2005fj,Byrnes:2007tm}
\begin{equation}
\label{local}
\mathcal{R}(\vec{x}) = \mathcal{R}_\mathrm{G}(\vec{x}) + \fnl \(\mathcal{R}_\mathrm{G}^2(\vec{x})- \left\langle \mathcal{R}_\mathrm{G}^2(\vec{x}) \right \rangle \),
\end{equation}
where $\mathcal{R}_\mathrm{G}({\vec{x}})$ follows Gaussian statistics, and $\fnl$ represents the dimensionless non-Gaussian parameters. It is worth noting that the non-Gaussianity parameter $\fnl$ is related to the commonly used notation $f_{\mathrm{NL}}$ through the relation $\fnl \equiv 3/5 f_{\mathrm{NL}}$. The non-Gaussian contributions are incorporated by defining the effective power spectrum of curvature perturbations, ${P}^{\mathrm{NG}}_{\mathcal{R}}(k)$, as~\cite{Cai:2018dig}
\begin{equation}
\label{NGP}
{P}^{\mathrm{NG}}_{\mathcal{R}}={P}_{\mathcal{R}}(k)+\fnl^2 \int_{0}^{\infty}\!\!\!\mathrm{d}v\!\! \int_{|1-v|}^{1+v}\mathrm{d}u~\frac{{P}_{\mathcal{R}}(uk){P}_{\mathcal{R}}(vk)}{u^2 v^2}.
\end{equation}

In the conformal Newton gauge, the metric perturbations can be expressed as
\e
    ds^2 = a^2(\eta)\left\{-(1+2\phi)\mathrm{d}\eta^2+[(1-2\phi)\delta_{ij}+h_{ij}]\mathrm{d}x^i \mathrm{d}x^j\right\},
\q
where $\eta$ represents the conformal time, $\phi$ is the Newtonian potential, and $h_{ij}$ corresponds to the tensor mode of the metric perturbation in the transverse-traceless gauge. The equation of motion for $h_{ij}$ can be obtained by considering the perturbed Einstein equation up to the second order, namely
\e\label{eqh}
h_{i j}^{\prime \prime}+2 \mathcal{H} h_{i j}^{\prime}-\nabla^{2} h_{i j}=-4 \mathcal{T}_{i j}^{\ell m} S_{\ell m},
\q
where the prime denotes a derivative with respect to the conformal time $\eta$, $\mathcal{H}\equiv \frac{a'}{a}$ represents the conformal Hubble parameter, and $\mathcal{T}_{i j}^{\ell m}$ is the transverse traceless projection operator in Fourier space. The source term $S_{ij}$,  which is of second order in scalar perturbations, reads
    \begin{equation}
S_{i j}=3 \phi \partial_i \partial_j \phi-\frac{1}{\mathcal{H}}\left(\partial_i \phi^{\prime} \partial_j \phi+\partial_i \phi \partial_j \phi^{\prime}\right)-\frac{1}{\mathcal{H}^2} \partial_i \phi^{\prime} \partial_j \phi^{\prime}.
\end{equation}
The characterization of SGWBs often involves describing their energy density per logarithmic frequency interval relative to the critical density $\rho_{c}(\eta)$,
\begin{equation}
\Omega_{\mathrm{GW}}(k, \eta) \equiv \frac{1}{\rho_c(\eta)} \frac{\mathrm{d} \rho_{\mathrm{GW}}(k, \eta)}{\mathrm{d} \ln k}=\frac{k^3}{48 \pi^2}\left(\frac{k}{\mathcal{H}}\right)^2 \overline{\left\langle\left|h_{\bm{k}}(\eta)\right|^2\right\rangle},
\end{equation}
where the overline represents an average over a few wavelengths. During the radiation-dominated era, GWs are generated by curvature perturbations, and their density parameter at the matter-radiation equality is denoted as $\ogw(k)=\ogw(k,\eta\rightarrow\infty)$. Using the relation between curvature perturbations $\mathcal{R}$ and scalar perturbations $\phi$ in the radiation-dominated era, $\phi=(2/3)\mathcal{R}$, we can calculate $\ogw(k)$ as~\cite{Kohri:2018awv}

\begin{equation}
\label{Omega_1}
\Omega_{\mathrm{GW}}(k) 
=\int_0^{\infty} \mathrm{d} v \int_{|1-v|}^{1+v} \mathrm{d} u \mathcal{T}\, {P}^{\mathrm{NG}}_{\mathcal{R}}(v k) {P}^{\mathrm{NG}}_{\mathcal{R}}(u k),
\end{equation} 
where the transfer function $\mathcal{T}=\mathcal{T}(u,v)$ is given by
 \begin{equation}
\begin{aligned}
\mathcal{T}(u,v)= & \frac{3}{1024 v^8 u^8}\left[4 v^2-\left(v^2-u^2+1\right)^2\right]^2\left(v^2+u^2-3\right)^2 \\
& \times\Bigg\{\left[\left(v^2+u^2-3\right) \ln \left(\left|\frac{3-(v+u)^2}{3-(v-u)^2}\right|\right)-4 v u\right]^2 \\
& +\pi^2\left(v^2+u^2-3\right)^2 \Theta(v+u-\sqrt{3})\Bigg\}.
\end{aligned}
\end{equation}
According to the~\Eq{NGP} and~\Eq{Omega_1}, $\Omega_{\mathrm{GW}}(k)$ can be expanded as 
\begin{equation}\label{ogwexpand}
    \Omega_{\mathrm{GW}}(k)= A^{2} \Omega^{(0)}(k) + A^{3} F_\text{NL}^{2} \Omega^{(2)}(k) + A^{4} F_\text{NL}^{4} \Omega^{(4)}(k),
\end{equation}
where $\Omega^{(0)}(k)$, $\Omega^{(2)}(k)$, and $\Omega^{(4)}(k)$ represent the corresponding integral terms, and $A \equiv \int  {P}_{\mathcal{R}}\,  \mathrm{d}\ln k$ is the amplitude of ${P}_{\mathcal{R}}$.
From~\Eq{ogwexpand}, we see that positive and negative $\fnl$ will generate identical SIGWs. In other words, positive and negative $\fnl$ are degenerate regarding their impact on SIGWs.

Using the relation between the wavenumber and frequency, $k=2\pi f$, we obtain the energy density fraction spectrum of SIGWs at the present time, 
\begin{equation}
\Omega_{\mathrm{GW}, 0}(f)=\Omega_{\mathrm{r}, 0}\left[\frac{g_{*, r}(T)}{g_{*, r}\left(T_{\mathrm{eq}}\right)}\right]\left[\frac{g_{*, s}\left(T_{\mathrm{eq}}\right)}{g_{*, s}(T)}\right]^{\frac{4}{3}} \Omega_{\mathrm{GW}}(k).
\end{equation}
It is given by the product of $\Omega_{\mathrm{GW}}(k)$, the present energy density fraction of radiation, $\Omega_{r,0}$, and two factors involving the effective degrees of freedom for entropy density, $g_{,s}$, and radiation, $g_{,r}$. To demonstrate the method, we adopt a commonly used power spectrum for ${P}_{\mathcal{R}}$, taking the lognormal form~\cite{Kohri:2018awv, Ferrante:2022mui}
\begin{equation}
{P}_{\mathcal{R}}(k) = \frac{A}{\sqrt{2\pi} \Delta} \exp\left(-\frac{\ln^2(k/k_*)}{2\Delta^2}\right),
\end{equation}
where $A$ is the amplitude, $k_*$ is the characteristic scale, and $\Delta$ denotes the width of the spectrum.

\begin{figure*}[htbp!]
	\centering
	\includegraphics[width=0.8\textwidth]{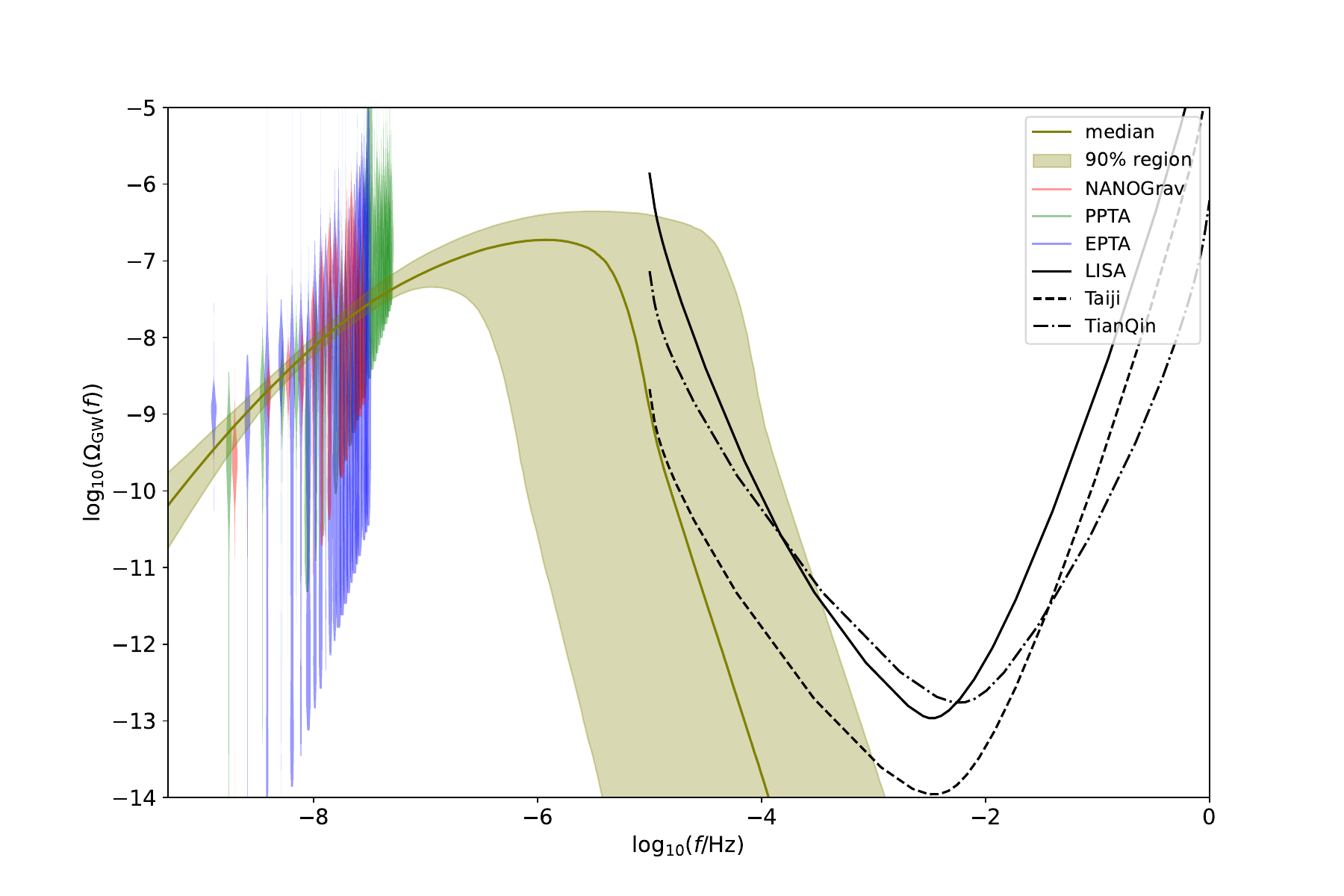}
	\caption{\label{ogw}The posterior predictive distribution for the energy density from SIGWs for the $\mathcal{M}_{\mathrm{NG}}$ model. The solid olive line is the median value, while the shaded region represents the $90\%$ credible region. We also show the energy density spectra derived from the free spectrum from NANOGrav 15-yr data set (red violins), PPTA DR3 (green violins), and EPTA DR2 (blue violins). The black solid, dashed, and dash-dotted lines represent the power-law integrated sensitivity curves for LISA, Taiji, and TianQin, respectively.}
\end{figure*}

\begin{table*}
    \centering
	\begin{tabular}{c|cccc}
		\hline\hline
		Parameter & $A$ & $\Delta$ & $f_*/\mathrm{Hz}$ & $|F_{\mathrm{NL}}|$\\[1pt]
		\hline
		 Prior& \quad log-$\mU(-3, 2)$\quad & \quad$\mU(0.05, 5)$\quad & \quad log-$\mU(-9, -2)$\qquad&  \quad log-$\mU(-5, 3)$\quad\\[1pt]
		Result for $\mathcal{M}_{\mathrm{G}}$ & $1.73^{+5.57}_{-1.47}$ & $3.24^{+0.70}_{-1.34}$ & \quad $3.25^{+51.1}_{-3.22} \times 10^{-5}$ & --\\[1pt]
		Result for $\mathcal{M}_{\mathrm{NG}}$ &$1.06^{+5.20}_{-1.02}$ & $3.36^{+1.10}_{-1.29}$ & \quad $1.81^{+45.3}_{-1.79} \times 10^{-5}$ & $\lesssim 13.9$\\[1pt]
  \hline
	\end{tabular}
	\caption{\label{tab:priors}Prior distributions and results for the model parameters. We consider two cases: a model with non-Gaussianity, $\mathcal{M}_{\mathrm{NG}}$, and a model without non-Gaussianity, $\mathcal{M}_{\mathrm{G}}$. Here $\mU$ and log-$\mU$ denote the uniform and log-uniform distributions, respectively. We quote each parameter's median value and $90\%$ equal-tail credible interval.}
\end{table*}

We note that a positive value of $\fnl$ will increase the abundance of PBHs for a given power spectrum of curvature perturbations. Conversely, a negative value of $\fnl$ will decrease the abundance of PBHs. This behavior highlights the impact of non-Gaussianity, quantified by $\fnl$, on the formation and abundance of PBHs. The Gaussian curvature perturbation $\mathcal{R}_\mathrm{G}$ can be determined by solving Eq.~\eqref{local} as~\cite{Byrnes:2012yx, Young:2013oia}
\begin{equation}
    \mathcal{R}_{\mathrm{G}}^\pm (\mathcal{R}) = \frac{1}{2 F_\text{NL}} \left( -1 \pm \sqrt{1+4 F_\text{NL} \mathcal{R}+4 F_\text{NL}^{2} \langle \mathcal{R}_{\mathrm{G}}^{2} \rangle} \right).
\end{equation}
PBHs are expected to form when the curvature perturbation exceeds a certain threshold value~$\mathcal{R}_{\mathrm{c}} \sim 1$~\cite{Musco:2004ak,Musco:2008hv,Musco:2012au,Harada:2013epa}. The PBH mass fraction at formation time can be calculated as~\cite{Young:2013oia}
\begin{equation}
\label{beta}
\beta(M) \simeq \frac{1}{2}\! \left\{\begin{array}{cl}
\!\!\!\operatorname{erfc}\!\!\left(\frac{\mathcal{R}_{\mathrm{G}}^+\left(\mathcal{R}_{\mathrm{c}}\right)}{\sqrt{2 \left\langle \mathcal{R}_\mathrm{G}^2 \right \rangle}}\right)+\operatorname{erfc}\!\!\left(-\frac{\mathcal{R}_{\mathrm{G}}^-\left(\mathcal{R}_{\mathrm{c}}\right)}{\sqrt{2 \left\langle \mathcal{R}_\mathrm{G}^2 \right \rangle}}\right) ; & F_{\mathrm{NL}}>0, \\
\!\!\operatorname{erf}\!\!\left(\frac{\mathcal{R}_{\mathrm{G}}^+\left(\mathcal{R}_{\mathrm{c}}\right)}{\sqrt{2 \left\langle \mathcal{R}_\mathrm{G}^2 \right \rangle}}\right) - \operatorname{erf}\!\!\left(\frac{\mathcal{R}_{\mathrm{G}}^-\left(\mathcal{R}_{\mathrm{c}}\right)}{\sqrt{2 \left\langle \mathcal{R}_\mathrm{G}^2 \right \rangle}}\right) ; & F_{\mathrm{NL}}<0.
\end{array}\right.
\end{equation}
One can define the total abundance of PBHs in the dark matter at present as~\cite{Sasaki:2018dmp}

\begin{equation}
\begin{aligned}
\label{fpbh}
f_{\mathrm{PBH}} &\equiv \frac{\Omega_{\mathrm{PBH}}}{\Omega_{\mathrm{CDM}}}=2.7 \times 10^8 \int_{-\infty}^{\infty} \mathrm{d} \ln M \\ & \times \left(\frac{g_{*,r}}{10.75}\right)^{3/ 4} \left(\frac{g_{*,s}}{10.75}\right)^{-1} 
 \left(\frac{M}{M_{\odot}}\right)^{-1 / 2} \beta(M),
\end{aligned}
\end{equation}
where $\Omega_{\mathrm{CDM}}$ is the cold dark matter density. 

\textbf{Data analyses and results.} We jointly use the NANOGrav 15-yr data set, PPTA DR3, and EPTA DR2 to estimate the model parameters. The ongoing efforts of these PTAs have lasted for more than a decade. Specifically, the NANOGrav 15-yr data set contains observations of $68$ pulsars with a time span of $16.03$ years~\cite{NANOGrav:2023hde}, PPTA DR3 contains observations of $32$ pulsars with a time span of up to $18$ years~\cite{Zic:2023gta}, and EPTA DR2 contains observations of $25$ pulsars with a time span of $24.7$ years~\cite{Antoniadis:2023lym}. These PTA data sets all present a stochastic signal consistent with the Hellings-Downs spatial correlations expected for an SGWB. If this signal is indeed of GW origin, it should share the same properties among these PTAs. Therefore, we combine the observations from these PTAs to estimate model parameters to increase the precision rather than using each individual PTA. In this letter, we use the free spectrum amplitude derived by each PTA with Hellings-Downs correlations. Given the time span $T_{\mathrm{obs}}$ of a PTA, the free spectrum starts with the lowest frequency $1/T_{\mathrm{obs}}$. NANOGrav, PPTA, and EPTA use $14$, $28$, and $24$ frequency components in their SGWB searches, respectively. Combining these data together results in $66$ frequencies of a free spectrum ranging from $1.28$~nHz to $49.1$~nHz. A visualization of the data used in the analyses is shown in \Fig{ogw}. In this work, we also consider the constraint from baryon acoustic oscillation and CMB~\cite{Planck:2018vyg} for the integrated energy-density fraction that $\int_{k_{\min }}^{\infty} h^2 \Omega_{\mathrm{GW}, 0}(k)\, d\ln k\lesssim 2.9 \times 10^{-7}$~\cite{Clarke:2020bil}, where $h = H_0 / ( 100 \mathrm{km}\, \mathrm{s}^{-1}\, \mathrm{Mpc}^{-1})=0.674$~\cite{Planck:2018vyg} is the dimensionless Hubble constant.

We use the time delay data released by each PTA. The time delay $d(f)$ can be converted to the power spectrum $S(f)$ by
\begin{equation}
d(f)=\sqrt{S(f) / T_{\mathrm{obs}}}.
\end{equation}
We then convert $S(f)$ to the characteristic strain, $h_c(f)$, by
\begin{equation}
h_c^2(f)=12 \pi^2 f^3 S(f).
\end{equation}
Further, we obtain the free spectrum energy density as
\begin{equation}
\hat{\Omega}_{\mathrm{GW}}(f)=\frac{2 \pi^2}{3 H_0^2} f^2 h_c^2(f) = \frac{8\pi^4}{H_0^2} T_{\mathrm{obs}} f^5 d^2(f).
\end{equation}
For each frequency $f_i$, with the posteriors of $\hat{\Omega}_{\mathrm{GW}}(f_i)$ at hand, we can estimate the corresponding kernel density $\mathcal{L}_i$. Therefore, the total likelihood is 
\e 
\mathcal{L}(\Lambda) = \prod_{i=1}^{66}  \mathcal{L}_i(\Omega_{\mathrm{GW}}(f_i, \Lambda)),
\q
where $\Lambda\equiv \{A, \Delta, f_*, |\fnl|\}$ is the collection of the model parameters.
We use \texttt{dynesty}~\cite{Speagle:2019ivv} sampler wrapped in \texttt{Bilby}~\cite{Ashton:2018jfp,Romero-Shaw:2020owr} package to search over the parameter space. The model parameters and their priors are summarized in \Table{tab:priors}.

\begin{figure}[htbp!]
	\centering
	\includegraphics[width=0.5\textwidth]{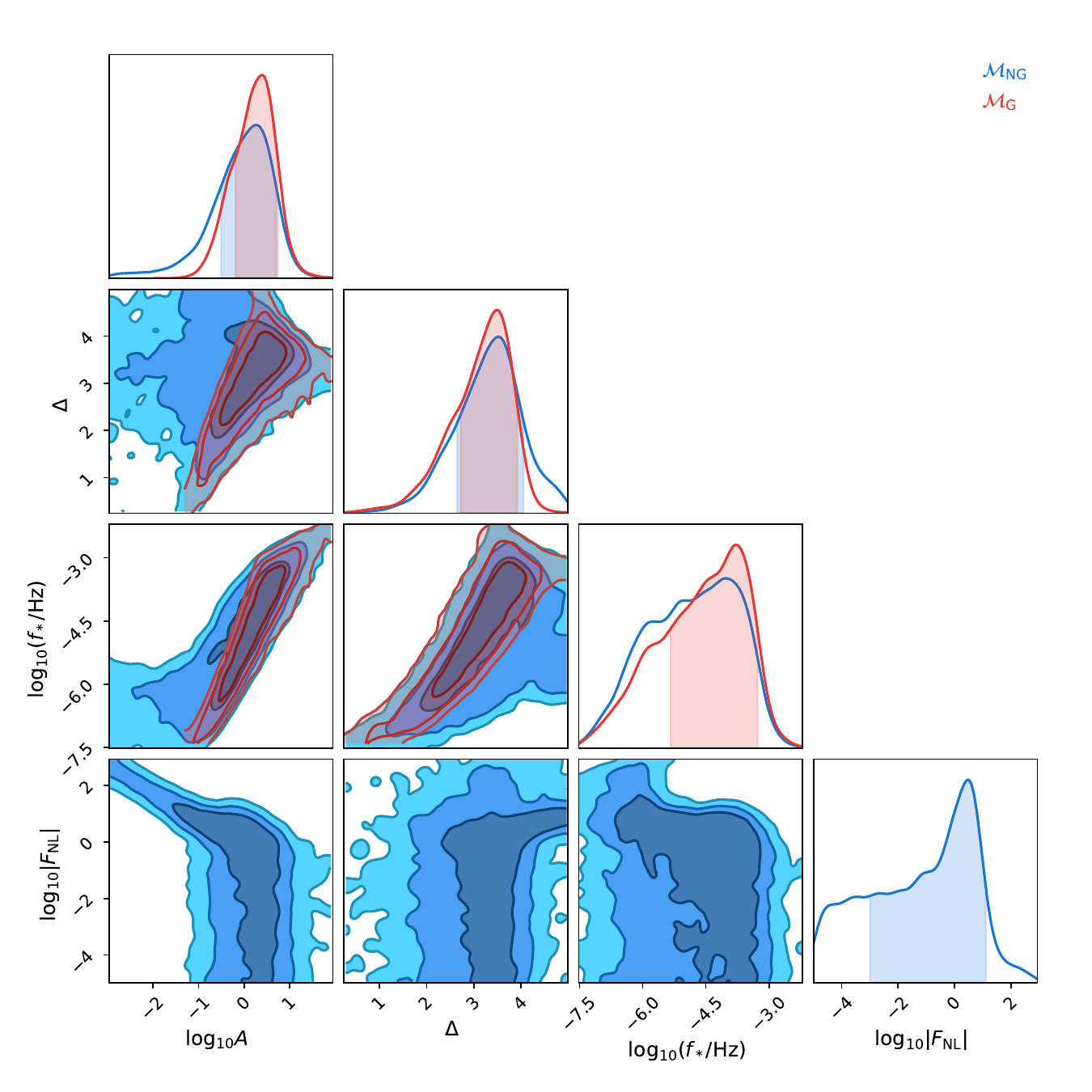}
	\caption{\label{post_All}One and two-dimensional marginalized posteriors of the parameters for the $\mathcal{M}_{\mathrm{G}}$ (red) model and the $\mathcal{M}_{\mathrm{NG}}$ (blue) model. We jointly use the PTA data from the NANOGrav 15-yr data set, PPTA DR3, and EPTA DR2. The contours in the two-dimensional plot correspond to the $1 \sigma$, $2 \sigma$, and $3 \sigma$ credible regions, respectively.}
\end{figure}

We consider two models: one without non-Gaussianity, $\mathcal{M}_{\mathrm{G}}$, and another with non-Gaussianity, $\mathcal{M}_{\mathrm{NG}}$. The posterior distributions for the parameters are shown in \Fig{post_All}, and the median and $90\%$ credible interval values for each parameter are summarized in \Table{tab:priors}. We note that the $\mathcal{M}_{\mathrm{G}}$ model has been studied by NANOGrav with their 15-yr data set, which is called \texttt{SIGW-GAUSS} in their paper. While we obtain consistent results, the combined data from NANOGrav, PPTA, and EPTA can constrain the parameters to higher precision than using the NANOGrav data set alone, as expected. For the $\MNG$ model, the $\fnl$ and $A$ parameters are generally degenerate. 
The combined data can constrain the amplitude to be $A = 1.06^{+5.20}_{-1.02}$, therefore constraining $|\fnl| \lesssim 13.9$. Since positive and negative $\fnl$ values are degenerate, we have $-13.9 \lesssim \fnl \lesssim 13.9$. 
Moreover, the abundance of PBHs cannot exceed that of dark matter, i.e., $f_{\mathrm{PBH} }\lesssim 1$. Using Eqs.~\eqref{beta} and \eqref{fpbh}, this limitation allows us to break the degeneracy and obtain constraints on $\fnl$ as $-13.9 \lesssim \fnl \lesssim -0.1$. 

\textbf{Summary and discussion.}
While the CMB and large-scale structure observations have provided increasingly precise measurements on the largest scales of the Universe, our knowledge of small scales remains limited, except for the constraints imposed by PBHs. PTAs, on the other hand, are an invaluable tool to probe the small-scale non-Gaussianity through SIGWs. Assuming the stochastic signal detected by the PTA collaborations originates from SIGWs, we jointly use the NANOGrav 15-yr data set, PPTA DR3, and EPTA DR2 to constrain the SIGWs accounting for non-Gaussianity. For the first time, we constrain the non-linear parameter as $|F_\mathrm{NL}|\lesssim 13.9$ for a lognormal power spectrum of the curvature perturbation. Furthermore, we obtain $-13.9 \lesssim F_\mathrm{NL}\lesssim -0.1$ to avoid overproduction of PBHs. Although we have only dealt with the lognormal power spectrum of curvature perturbations, the method and the framework proposed in this work can be easily extended to different types of power spectra. For instance, a similar constraint on the non-Gaussianity parameter associated with the broken power-law spectrum is presented in the \textit{Supplementary Material}.

The constraints on primordial non-Gaussianity of local type have significant implications for inflation models that involve scalar fields, other than the inflaton, in generating the primordial curvature perturbations. For instance, adiabatic curvaton models predict that~\cite{Bartolo:2003jx,Bartolo:2010qu}
\begin{equation}
f_\mathrm{NL} = \frac{5}{3} F_\mathrm{NL} =\frac{5}{4 r_\mathrm{D}} - \frac{5 r_\mathrm{D}}{6}-\frac{5}{3},
\end{equation}
when the curvaton field has a quadratic potential~\cite{Lyth:2001nq,Lyth:2002my,Lyth:2005fi,Malik:2006pm,Sasaki:2006kq}. 
Here the parameter $r_\mathrm{D} = 3\rho_\mathrm{curvaton}/(3\rho_\mathrm{curvaton} + 4 \rho_\mathrm{radiation})$ 
represents the ``curvaton decay fraction" at the time of curvaton decay under sudden decay approximation. 
Our constraint $|F_\mathrm{NL}|\lesssim 13.9$ implies 
\begin{equation}
    r_{\mathrm{D}} \gtrsim 0.05 \quad (95\%),
\end{equation}
and the further constraint that $F_\mathrm{NL} \lesssim -0.1$ yields
\begin{equation}
    r_\mathrm{D} \gtrsim 0.62 \quad (95\%),
\end{equation}
indicating that the curvaton field has a non-negligible energy density when it decays. Our findings, therefore, pave the way to constrain inflation models with PTA data.

Furthermore, as indicated in \Fig{ogw}, the energy density spectrum of SIGW can generally be extended to the frequency band of the space-borne GW detector. Therefore, the multi-band observations of PTAs with the forthcoming space-borne GW detectors, such as LISA/Taiji/TianQin, will provide a complementary investigation of non-Gaussianity.

\textbf{Note added.} While finalizing this work, we found two parallel independent studies~\cite{Franciolini:2023pbf,Wang:2023ost} that also explore the potential connection between the NANOGrav signal and SIGWs associated with non-Gaussianity. In particular, Ref.~\cite{Franciolini:2023pbf} focused on the significance of the non-Gaussianity parameter to address the issue of PBH overproduction; however, it did not constrain the non-Gaussianity parameter with PTA data. On the other hand, Ref.~\cite{Wang:2023ost} did obtain a constraint on the non-Gaussianity parameter with the NANOGrav data, but it is noteworthy that Ref.~\cite{Wang:2023ost} adopted a less rigorous approach by manually fixing other model parameters. In contrast, our approach employs a comprehensive Bayesian inference methodology, providing a rigorous constraint on the non-Gaussianity parameter using multiple PTA data sets.

\emph{Acknowledgments.}
The contour plots are generated with \texttt{ChainConsumer}~\cite{Hinton2016}.
LL is supported by the National Natural Science Foundation of China (Grant No.~12247112 and No.~12247176) and the China Postdoctoral Science Foundation Fellowship No.~2023M730300. ZCC is supported by the National Natural Science Foundation of China (Grant No.~12247176 and No.~12247112) and the China Postdoctoral Science Foundation Fellowship No.~2022M710429. QGH is supported by grants from NSFC (Grant No.~12250010, 11975019, 11991052, 12047503), Key Research Program of Frontier Sciences, CAS, Grant No.~ZDBS-LY-7009, CAS Project for Young Scientists in Basic Research YSBR-006, the Key Research Program of the Chinese Academy of Sciences (Grant No.~XDPB15). 
\bibliography{ref}

\clearpage
\newpage
\maketitle
\onecolumngrid
\begin{center}
\textbf{\large \papertitle} 
\\ 
\vspace{0.05in}
{Lang Liu, Zu-Cheng Chen, and Qing-Guo Huang}
\\ 
\vspace{0.05in}
{ \it Supplementary Material}
\end{center}
\onecolumngrid
\setcounter{equation}{0}
\setcounter{figure}{0}
\setcounter{section}{0}
\setcounter{table}{0}
\setcounter{page}{1}
\makeatletter
\renewcommand{\theequation}{S\arabic{equation}}
\renewcommand{\thefigure}{S\arabic{figure}}
\renewcommand{\thetable}{S\arabic{table}}

\vspace{-0.5cm}
\section{\label{supp:1} Impact of non-Gaussianity on PBH abundance}

\begin{figure}[htbp!]
	\centering
	\includegraphics[width=0.6\textwidth]{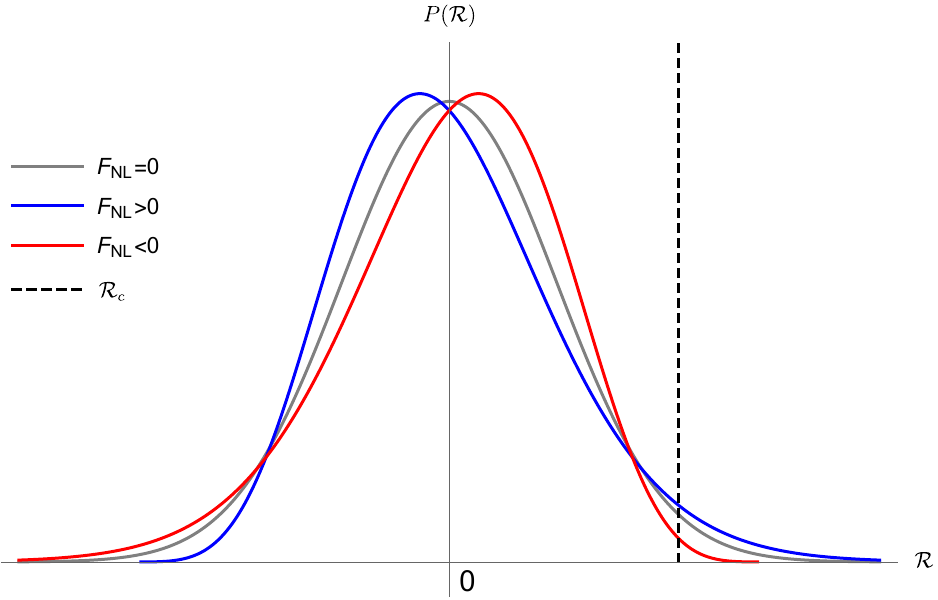}
	\caption{\label{PR} An illustration of the impact of the non-Gaussianity parameter $\fnl$ on the probability distribution of the curvature perturbation $\mathcal{R}$. Notably, non-Gaussianity induces a skew in the distribution. 
 A positive $\fnl$ extends the tail of the probability distribution for $\mathcal{R}>0$, thereby elevating the likelihood of $\mathcal{R} > \mathcal{R}_c$ and consequently leading to increased production of PBHs. Conversely, a negative $\fnl$ diminishes the PBH production.}
\end{figure}

We consider the model of local quadratic non-Gaussianity as~\cite{Verde:1999ij,Verde:2000vr,Komatsu:2001rj,Bartolo:2004if,Boubekeur:2005fj,Byrnes:2007tm}
\begin{equation}
\label{local-s}
\mathcal{R}(\vec{x}) = \mathcal{R}_\mathrm{G}(\vec{x}) + \fnl \(\mathcal{R}_\mathrm{G}^2(\vec{x})- \left\langle \mathcal{R}_\mathrm{G}^2(\vec{x}) \right \rangle \),
\end{equation}
where $\mathcal{R}_\mathrm{G}({\vec{x}})$ follows Gaussian statistics and the term $\left\langle \mathcal{R}_\mathrm{G}^2 \right \rangle$ is included to ensure that the expectation value for the curvature perturbations remains zero, $\langle\mathcal{R}\rangle=0$. Solving this equation to find $\mathcal{R}_{G}$ as a function of $\mathcal{R}$ gives two solutions~\cite{Byrnes:2012yx, Young:2013oia}
\begin{equation}
\label{bound}
    \mathcal{R}_{\mathrm{G}}^\pm (\mathcal{R}) = \frac{1}{2 F_\text{NL}} \left( -1 \pm \sqrt{1+4 F_\text{NL} \mathcal{R}+4 F_\text{NL}^{2} \langle \mathcal{R}_{\mathrm{G}}^{2} \rangle} \right).
\end{equation}
Making a formal change of variable, we can achieve the non-Gaussian probability distribution function as 
\begin{equation}
P_{\mathrm{NG}}(\mathcal{R})d\mathcal{R}=\sum_{i=\pm}\left|\frac{d\mathcal{R}_{\mathrm{G}}^{i}(\mathcal{R})}{d\mathcal{R}}\right|P_{\mathrm{G}}\left(\mathcal{R}_{\mathrm{G}}^{i}(\mathcal{R})\right)d\mathcal{R},
\label{changeofvariable}
\end{equation}
where 
\begin{equation}
\label{gaussian}
P_{\mathrm{G}}(\mathcal{R})=\frac{1}{\sqrt{2\pi \left\langle \mathcal{R}_\mathrm{G}^2(\vec{x}) \right \rangle}}\exp\left(-\frac{1}{2}\frac{\mathcal{R}^{2}}{\left\langle \mathcal{R}_\mathrm{G}^2(\vec{x}) \right \rangle}\right)
\end{equation}
is the Gaussian probability distribution function. The PBH mass fraction at formation time is given by 
\begin{equation}
\label{beta-ng}
\beta\simeq\int_{\mathcal{R}_c}^{\mathcal{R}_{\rm max}}P_{\rm NG}(\mathcal{R})d\mathcal{R}.
\end{equation}
When $F_{\rm NL}$ is positive (or zero), $\mathcal{R}_{\rm max}=\infty$. However, if $F_{\rm NL}$ is negative, then $\mathcal{R}$ is bounded from Eq. \eqref{bound}. In this case, $\mathcal{R}_{\rm max}$ is given by the expression \cite{Young:2013oia}
\begin{equation}
\mathcal{R}_{\rm max}=-\frac{1}{4F_{\rm NL}}\left(1+4F_{\rm NL}^{2}\langle \mathcal{R}_{\mathrm{G}}^{2} \rangle\right).
\end{equation}
\Fig{PR} illustrates the impact of the non-Gaussianity parameter $\fnl$ on the probability distribution of the curvature perturbation $\mathcal{R}$.
The primary effect of $F_{\rm NL}$ is to skew the distribution of $\mathcal{R}$. When $F_{\rm NL}$ is positive, we observe a peak in the distribution for negative values of $\mathcal{R}$, along with a large tail for positive values of $\mathcal{R}$. Conversely, for negative $F_{\rm NL}$, we observe a peak in the distribution for positive values of $\mathcal{R}$, along with a small tail for positive values of $\mathcal{R}$. If the maximum curvature perturbation $\mathcal{R}_{\rm max}$ satisfies $\mathcal{R}_{\rm max}<\mathcal{R}_c$, then no PBH formation would be expected for these values.  In summary, a positive value of $\fnl$ will increase the abundance of PBHs, while a negative value of $\fnl$ will decrease the abundance of PBHs. From Eq. \eqref{beta-ng}, for $F_{\rm NL}>0$, we have 
\begin{equation}
\beta\simeq \frac{1}{2} \left[ \operatorname{erfc}\!\!\left(\frac{\mathcal{R}_{\mathrm{G}}^+\left(\mathcal{R}_{\mathrm{c}}\right)}{\sqrt{2 \left\langle \mathcal{R}_\mathrm{G}^2 \right \rangle}}\right)+\operatorname{erfc}\!\!\left(-\frac{\mathcal{R}_{\mathrm{G}}^-\left(\mathcal{R}_{\mathrm{c}}\right)}{\sqrt{2 \left\langle \mathcal{R}_\mathrm{G}^2 \right \rangle}}\right) \right],
\end{equation}
and $F_{\rm NL}<0$, we have 
\begin{equation}
\beta \simeq \frac{1}{2} \left[\operatorname{erf}\!\!\left(\frac{\mathcal{R}_{\mathrm{G}}^+\left(\mathcal{R}_{\mathrm{c}}\right)}{\sqrt{2 \left\langle \mathcal{R}_\mathrm{G}^2 \right \rangle}}\right) - \operatorname{erf}\!\!\left(\frac{\mathcal{R}_{\mathrm{G}}^-\left(\mathcal{R}_{\mathrm{c}}\right)}{\sqrt{2 \left\langle \mathcal{R}_\mathrm{G}^2 \right \rangle}}\right) \right].
\end{equation}

\section{\label{supp:bpl}Broken power-law spectrum}
Here, we consider another curvature perturbation spectrum characterized by a \emph{broken power-law} form.
The broken power-law spectrum can describe a common class of spectra that exhibit peaks, often associated with single-field inflation or curvaton models. The functional form is given by
\e\label{eq:PPL}
    \mathcal{P}^{\rm BPL}_{\mathcal{R}}(k)
    =A \frac{\left(\alpha+\beta\right)^{\gamma}}{\left[\beta\left(k / k_*\right)^{-\alpha/\gamma}+\alpha\left(k / k_*\right)^{\beta/\gamma}\right]^{\gamma}}.
\q
Here, $\alpha$ and $\beta$ are both positive numbers that describe the growth and decay of the spectrum around the peak, respectively. 
The parameter $\gamma$ characterizes the flatness of the peak, with typical values for $\alpha$ being $\alpha \lesssim 4$~\cite{Byrnes:2018txb}.
Additionally, in quasi-inflection-point models that give rise to stellar-mass PBHs, one typically expects $\beta \gtrsim 0.5$, while for curvaton models $\beta \gtrsim 2$.

\begin{table}[thbp!]
    \centering
	\begin{tabular}{c|cccccc}
		\hline\hline
		Parameter & $f_*/\mathrm{Hz}$ &  $A$ & $\alpha$\quad & $\beta$ & $\gamma$ & $|F_{\mathrm{NL}}|$\\[1pt]
		\hline
		 Prior& log-$\mU(-9, -2)$ &log-$\mU(-3, 2)$ & $\mU(0, 2)$\quad &  $\mU(0, 10)$\quad & $\mU(0, 10)$\quad & log-$\mU(-5, 4)$\\[1pt]
		Result for $\mathcal{M}_{\mathrm{G}}$ & $3.2^{+12.3}_{-2.5}\times 10^{-7}$ & $0.82^{+0.58}_{-0.37}$ & $0.86^{+0.76}_{-0.21}$ & $4.6^{+4.8}_{-3.60}$ & $4.9^{+4.6}_{-4.2}$ & --\\[1pt]
		Result for $\mathcal{M}_{\mathrm{NG}}$ & $2.7^{+9.8}_{-2.1}\times 10^{-7}$ & $0.67^{+0.67}_{-0.65}$ & $0.83^{+0.79}_{-0.30}$ & $4.8^{+4.50}_{-3.7}$ & $5.1^{+4.4}_{-4.4}$ & $\lesssim 11.1$\\[1pt]
  \hline
	\end{tabular}
	\caption{\label{tab:priors_bpl}Same as \Table{tab:priors}, but for the broken power-law spectrum.}
\end{table}

\begin{figure}[htbp!]
	\centering
	\includegraphics[width=0.9\textwidth]{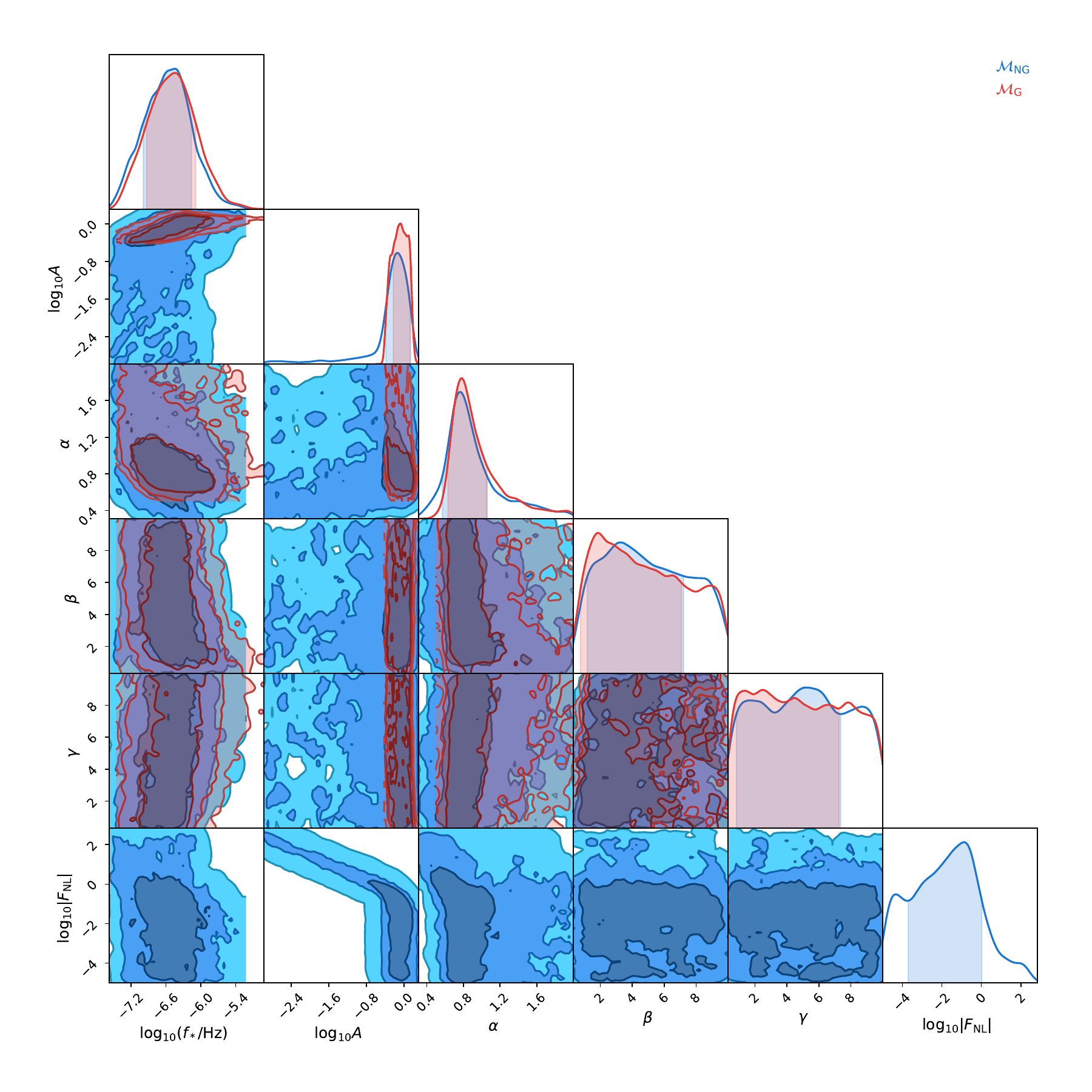}
	\caption{\label{post_All_bpl}Same as \Fig{post_All}, but for the broken power-law spectrum.}
\end{figure}

The model parameters, along with their priors and inferred values are summarized in \Table{tab:priors_bpl}, while the posterior distributions for the parameters are shown in \Fig{post_All_bpl}.
The joint analysis, incorporating the NANOGrav 15-yr data set, PPTA DR3, and EPTA DR2, yields the constraint: $-11.1 \lesssim \fnl \lesssim 11.1$. 
To avoid the PBH overproduction, we further constrain the non-Guassianity parameter as $-11.1 \lesssim \fnl \lesssim -0.3$.

\end{document}